\documentclass[12pt]{article}

\usepackage{amsmath}
\usepackage{graphicx}
\usepackage{cite}

\newcommand{\ket}[1]{| #1 \rangle}
\newcommand{\bra}[1]{\langle #1 |}
\newcommand{\hcs}[1]{#1^\dagger #1}
\newcommand{\expv}[1]{\langle #1 \rangle}
\begin{document}

\begin{center}
{\Large\bf Information and fidelity
in projective measurements}
\vskip .6 cm
Hiroaki Terashima
\vskip .4 cm
{\it Department of Physics, Faculty of Education, Gunma University, \\
Maebashi, Gunma 371-8510, Japan}
\vskip .6 cm
\end{center}

\begin{abstract}
In this study, we explicitly calculate information and fidelity
of an $r$-rank projective measurement on a completely unknown state
in a $d$-dimensional Hilbert space.
We also show a tradeoff between information and fidelity
at the level of a single outcome
and discuss the efficiency of measurement
with respect to fidelity.
\end{abstract}

\begin{flushleft}
{\footnotesize
}
\end{flushleft}

\section{Introduction}
In quantum theory, a measurement that provides information
about a physical system
inevitably changes the state of the system
depending on the outcome of the measurement.
This is an interesting property of quantum measurement
not only in the foundations of quantum mechanics
but also in quantum information processing
and communication~\cite{NieChu00}
such as quantum cryptography~\cite{BenBra84,Ekert91,Bennet92,BeBrMe92}.
Therefore, there have been many discussions regarding
the tradeoffs between information gain and state change
using various formulations~\cite{FucPer96,Banasz01,FucJac01,%
BanDev01,DArian03,Ozawa04,Maccon06,Sacchi06,BusSac06,BuHaHo07}.
For example, Banaszek~\cite{Banasz01} has shown
an inequality between mean estimation fidelity
and mean operation fidelity that quantifies
information gain and state change, respectively.

In connection with such tradeoffs,
the author~\cite{Terash10,Terash11} has recently
discussed tradeoffs together with
physical reversibility~\cite{UeImNa96,Ueda97} of measurement
in the context of reversibility
in quantum measurement~\cite{UedKit92,Imamog93,Royer94,%
MabZol96,NieCav97,KoaUed99,Ban01,TerUed05,KorJor06,%
TerUed07,KNABHL08,KCRK09,SuAlZu09,TerUed07b,XuZho10}.
In particular, the author~\cite{Terash11} has shown
tradeoffs among information gain,
state change, and physical reversibility
in the case of single-qubit measurements.
An important feature of these tradeoffs
is that they occur at the level of a single outcome
without averaging all possible
outcomes~\cite{FucPer96,Banasz01,BanDev01,Sacchi06}.
This feature originates from the fact that
the physical reversibility of measurements
suggests quantifying the information gain and
the state change for each single outcome,
because in physically reversible measurements,
a state recovery with information erasure
(see the Erratum of \cite{Royer94})
occurs because of the post-selection of outcomes.
However, the explicit calculations
in the previous studies~\cite{Terash10,Terash11}
were only performed with two-level systems or qubits.

In this study,
we calculate information gain and state change
in a projective measurement of rank $r$ on a $d$-level system
assumed to be in a completely unknown state.
We evaluate the amount of information gain
by a decrease in Shannon entropy~\cite{DArian03,TerUed07b}
and the degree of state change by fidelity~\cite{Uhlman76}
to express them as functions of $r$ and $d$.
These results lead to a tradeoff
between information gain and state change
at a single outcome level.
We also consider the efficiency of the measurement
with respect to fidelity.
Of course, projective measurements
are not physically reversible~\cite{UeImNa96}.
However, they would correspond to special points
as the most informative but the least reversible measurements
in the tradeoffs among information gain,
state change, and physical reversibility
in general measurements on a $d$-level system.

The rest of this paper is organized as follows:
Section~\ref{sec:formulation} explains the procedure to quantify
information gain and state change
and calculates them in the case of
an $r$-rank projective measurement on a $d$-level system.
Section~\ref{sec:tradeoff} discusses
a tradeoff between information gain and state change
and considers efficiency of the measurement
with respect to the state change.
Section~\ref{sec:conclude} summarizes our results.

\section{\label{sec:formulation}Information and Fidelity}

% Information

We evaluate the amount of information provided
by a quantum measurement as follows.
Suppose that the pre-measurement state of a system
is known to be one of
the predefined pure states $\{\ket{\psi(a)}\}$, $a=1,\ldots,N$,
with equal probability $p(a)=1/N$~\cite{TerUed07b,Terash10,Terash11},
although the index $a$ of the pre-measurement state is unknown to us.
Thus, the lack of information about the state of the system
can be evaluated by Shannon entropy as
\begin{equation}
  H_0=-\sum_a p(a)\log_2 p(a)=\log_2 N
\label{eq:h0}
\end{equation}
before measurement,
where we have used the Shannon entropy rather than
the von Neumann entropy of the mixed state
$\hat{\rho}=\sum_a p(a) \ket{\psi(a)}\bra{\psi(a)}$
because what we are uncertain about is the classical variable $a$
rather than the predefined quantum state $\ket{\psi(a)}$.
If the pre-measurement state is completely unknown,
as is usually the case in quantum measurement,
then the set of the predefined states, $\{\ket{\psi(a)}\}$, consists of
all possible pure states of the system with $N\to\infty$.
Each state can be expanded by an orthonormal basis $\{\ket{k}\}$ as
\begin{equation}
  \ket{\psi(a)}=\sum_k c_k(a)\, \ket{k}
\label{eq:basis}
\end{equation}
with $k=1,2,\ldots,d$, where
$d$ is the dimension of 
the Hilbert space associated with the system.
The coefficients $\{c_k(a)\}$ obey the
normalization condition
\begin{equation}
\sum_k \left| c_k(a)\right|^2=1.
\label{eq:norm}
\end{equation}

We next perform a quantum measurement on the system
to obtain the information about its state.
A quantum measurement is generally described by a set of
measurement operators $\{\hat{M}_m\}$~\cite{DavLew70,NieChu00}
that satisfies
\begin{equation}
\sum_m\hcs{\hat{M}_m}=\hat{I},
\label{eq:compcond}
\end{equation}
where $\hat{I}$ is the identity operator.
If the system to be measured is in a state $\ket{\psi}$,
the measurement yields an outcome $m$ with probability
\begin{equation}
 p_m=\bra{\psi}\hcs{\hat{M}_m}\ket{\psi}
\label{eq:probability}
\end{equation}
and then causes a state reduction of the measured system into
\begin{equation}
\ket{\psi_m}=\frac{1}{\sqrt{p_m}}\hat{M}_m\ket{\psi}.
\label{eq:reduction}
\end{equation}

Here we consider performing a projective measurement
because it is the most informative.
In particular,
we perform a measurement where the process
yielding a particular outcome $m$ is described by
a projection operator of rank $r$ ($r=1,2,\ldots,d$);
that is,
the measurement operator corresponding to the outcome $m$
is written without loss of generality as
\begin{equation}
 \hat{M}_m=\kappa_m\hat{P}^{(r)}=\kappa_m\sum_{k=1}^{r} \ket{k}\bra{k}
\label{eq:operator}
\end{equation}
by relabeling the orthonormal basis,
where $\kappa_m$ is a complex number.
The other measurement operators are irrelevant
as long as condition (\ref{eq:compcond}) is satisfied,
since our interest is only at the level of a single outcome.
The measurement then yields the outcome $m$ with probability
\begin{equation}
 p(m|a) 
     = |\kappa_m|^2  \sum_{k=1}^{r}\left| c_k(a)\right|^2
     \equiv |\kappa_m|^2  q_m(a)
\label{eq:prob}
\end{equation}
when the pre-measurement state is $\ket{\psi(a)}$
from Eqs.~(\ref{eq:basis}) and (\ref{eq:probability}).
Since the probability for $\ket{\psi(a)}$ is $p(a)=1/N$,
the total probability for the outcome $m$ becomes
\begin{equation}
 p(m) =\sum_a  p(m|a)\,p(a)=\frac{1}{N}\sum_a  |\kappa_m|^2  q_m(a)
      = |\kappa_m|^2 \overline{q_m},
\label{eq:totalprob}
\end{equation}
where the overline denotes the average over $a$,
\begin{equation}
   \overline{f} \equiv \frac{1}{N}\sum_a f(a).
\label{eq:overline}
\end{equation}
On the contrary,
Bayes' rule states that given the outcome $m$, the probability for
the pre-measurement state $\ket{\psi(a)}$ is given by
\begin{equation}
  p(a|m) =\frac{p(m|a)\,p(a)}{p(m)}=\frac{q_m(a)}{N\,\overline{q_m}}.
\label{eq:conditional}
\end{equation}
Thus, the lack of information about the pre-measurement state
can be evaluated by Shannon entropy as
\begin{equation}
  H(m) =-\sum_a p(a|m)\log_2 p(a|m)
\end{equation}
after the measurement yields the outcome $m$.
Therefore, we define information gain by the measurement with
the \emph{single} outcome $m$ as the decrease
in Shannon entropy~\cite{DArian03,TerUed07b}
\begin{equation}
  I(m) \equiv H_0-H(m)
    =\frac{\overline{q_m\log_2 q_m} -\overline{q_m}\log_2 \overline{q_m}}
     {\overline{q_m}},
\label{eq:info}
\end{equation}
which is free from
the divergent term $\log_2 N\to\infty$ in Eq.~(\ref{eq:h0})
owing to the assumption that
the probability distribution $p(a)$ is uniform.

\begin{figure}
\begin{center}
\includegraphics[scale=0.6]{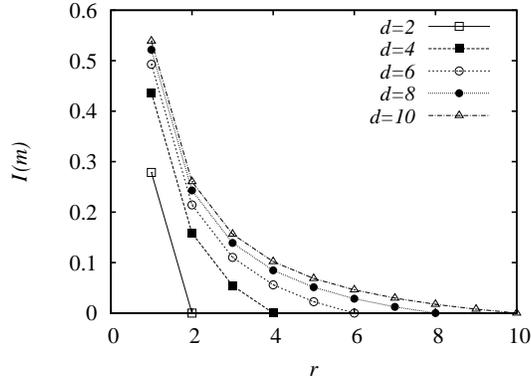}
\end{center}
\caption{\label{fig1}
Information gain $I(m)$
when the projective measurement yields the outcome $m$
as a function of rank $r$ for $d=2,4,6,8,10$.}
\end{figure}

In order to explicitly calculate the information gain (\ref{eq:info}),
we introduce parametrization of the coefficients $\{c_k(a)\}$.
Let $\alpha_k(a)$ and $\beta_k(a)$ be
the real and imaginary parts of $c_k(a)$, respectively:
\begin{equation}
  c_k(a)=\alpha_k(a)+i\beta_k(a).
\end{equation}
The normalization condition (\ref{eq:norm}) then becomes
\begin{equation}
   \sum_k \left[\alpha_k(a)^2+\beta_k(a)^2\right]=1.
\end{equation}
Note that this is the condition for a point to be on
the unit sphere in $2d$ dimensions.
This means that $\{\alpha_k(a)\}$ and $\{\beta_k(a)\}$
can be expressed by hyperspherical coordinates
$(\theta_1,\theta_2,\ldots,\theta_{2d-2},\phi)$ as~\cite{TerUed07b}
\begin{align}
   \alpha_1(a) &= \sin\theta_{2d-2}\sin\theta_{2d-3}\cdots
                   \sin\theta_3\sin\theta_2\sin\theta_1\cos\phi, \notag \\
   \beta_1(a) &= \sin\theta_{2d-2}\sin\theta_{2d-3}\cdots
                   \sin\theta_3\sin\theta_2\sin\theta_1\sin\phi, \notag  \\
   \alpha_2(a) &= \sin\theta_{2d-2}\sin\theta_{2d-3}\cdots
                   \sin\theta_3\sin\theta_2\cos\theta_1, \notag  \\
   \beta_2(a) &= \sin\theta_{2d-2}\sin\theta_{2d-3}\cdots
                   \sin\theta_3\cos\theta_2, \label{eq:parameter}  \\
             &\vdots  \notag \\
   \alpha_d(a) &= \sin\theta_{2d-2}\cos\theta_{2d-3}, \notag  \\
   \beta_d(a) &= \cos\theta_{2d-2},\notag 
\end{align}
where $0\le \phi < 2\pi$ and
$0\le \theta_p \le \pi$ with $p=1,2,\ldots,2d-2$.
The index $a$ now represents
the angles $(\theta_1,\theta_2,\ldots,\theta_{2d-2},\phi)$ and thus
the summation over $a$ is replaced
with an integral over the angles as
\begin{equation}
   \frac{1}{N}\sum_a   \quad\longrightarrow\quad
     \frac{(d-1)!}{2\pi^{d}}\int^{2\pi}_0 d\phi\,
     \prod^{2d-2}_{p=1}\int^\pi_0 d\theta_p\sin^p \theta_p.
\end{equation}
From Eqs.~(\ref{eq:prob}) and (\ref{eq:overline}), we get
\begin{equation}
 q_m(a)=\begin{cases}
           \displaystyle \prod^{2d-2}_{p=2r-1}\sin^2 \theta_p & (r<d) \\[6pt]
                 \quad\qquad 1     &(r=d)
        \end{cases}
\end{equation}
and
\begin{equation}
 \overline{q_m}=\frac{r}{d}
\label{eq:qbar}
\end{equation}
using the integral formula
\begin{equation}
  \int^\pi_0 d\theta \,\sin^n \theta =\sqrt{\pi}\,
  \frac{\Gamma\left(\frac{n+1}{2}\right)}
      {\Gamma\left(\frac{n+2}{2}\right)},
\end{equation}
where $n>-1$ with the Gamma function $\Gamma(n)$.
Similarly, using
\begin{equation}
  \int^{\pi}_0 d\theta \,\sin^n \theta\, \log_2\sin  \theta
    = \sqrt{\pi}\,\frac{\Gamma\left(\frac{n+1}{2}\right)}
        {\Gamma\left(\frac{n+2}{2}\right)}
     \left[(-1)^{n+1}+\sum_{k=1}^{n}\frac{(-1)^{n+k+1}}{k\ln2}\right]
\end{equation}
for $n>-1$~\cite{GraRyz07} with $\log_2 x=\ln x/\ln 2$, we obtain
\begin{equation}
 \overline{q_m \log_2 q_m}
        =-\frac{r}{d\ln2}\Bigl[\eta(d)- \eta(r)\Bigr],
\end{equation}
where
\begin{equation}
 \eta(n)\equiv \sum^{n}_{k=1}\frac{1}{k}.
\end{equation}
Therefore, the total probability (\ref{eq:totalprob})
and the information gain (\ref{eq:info}) are calculated
to be
\begin{equation}
 p(m) =|\kappa_m|^2 \frac{r}{d}
\end{equation}
and
\begin{equation}
  I(m)=\log_2\frac{d}{r}-\frac{1}{\ln2}\Bigl[\eta(d)- \eta(r)\Bigr],
\label{eq:information}
\end{equation}
respectively.
Figure~\ref{fig1} shows the information gain $I(m)$
as a function of rank $r$ for $d=2,4,6,8,10$.
As shown in Fig.~\ref{fig1},
the information gain monotonically decreases as $r$ increases
and becomes $0$ at $r=d$.
Note that when $r=d$, the measurement corresponds to
an uninformative identity operation,
since the measurement operator (\ref{eq:operator})
reduces to the identity operator $\hat{I}$
except for the constant $\kappa_m$.
In contrast,
when $r$ is fixed,
the information gain monotonically increases as $d$ increases.
Thus, taking the limit of Eq.~(\ref{eq:information})
as $d$ goes to infinity at $r=1$,
we find the upper bound on information gain as
\begin{equation}
I(m)\to \frac{1}{\ln2}\left(1-\gamma \right)\simeq0.610,
\end{equation}
where $\gamma$ is Euler's constant.

% Fidelity

\begin{figure}
\begin{center}
\includegraphics[scale=0.6]{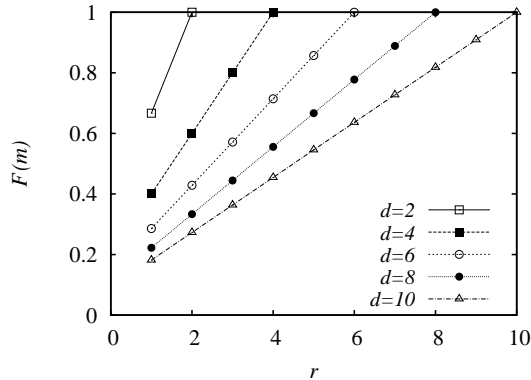}
\end{center}
\caption{\label{fig2}
Fidelity $F(m)$
when the projective measurement yields the outcome $m$
as a function of rank $r$ for $d=2,4,6,8,10$.}
\end{figure}

On the other hand,
the measurement  changes the state of the measured system.
When the pre-measurement state is $\ket{\psi(a)}$
and the measurement outcome is $m$,
the post-measurement state is
\begin{equation}
   \ket{\psi(m,a)} = \frac{1}{\sqrt{p(m|a)}}\,\kappa_m\hat{P}^{(r)}
       \ket{\psi(a)}
\end{equation}
according to Eqs.~(\ref{eq:reduction}) and (\ref{eq:operator}).
To quantify this state change,
we use fidelity~\cite{Uhlman76,NieChu00}
between the pre-measurement and post-measurement states, namely
\begin{equation}
   F(m,a) = \bigl|\expv{\psi(a)|\psi(m,a)}\bigr|=\sqrt{q_m(a)}.
\end{equation}
This fidelity decreases as
the measurement increasingly changes the state of the system.
Averaging it over $a$ with the probability (\ref{eq:conditional}),
we evaluate the degree of state change as
\begin{equation}
   F(m) =\sum_a p(a|m)\bigl[F(m,a)\bigr]^2
     = \frac{\overline{q_m^2}}{\;\overline{q_m}\;}
\label{eq:fide}
\end{equation}
after the measurement yields the outcome $m$,
where for simplicity, we have averaged the squared fidelity
rather than the fidelity.
The fidelity (\ref{eq:fide}) can be explicitly calculated
using the parameterization (\ref{eq:parameter}) as
\begin{equation}
   F(m)=\frac{r+1}{d+1},
\label{eq:fidelity}
\end{equation}
because of Eq.~(\ref{eq:qbar}) and
\begin{equation}
   \overline{q_m^2}=\frac{r(r+1)}{d(d+1)}.
\end{equation}
Figure~\ref{fig2} shows the fidelity $F(m)$
as a function of rank $r$ for $d=2,4,6,8,10$.
In contrast to information gain,
fidelity monotonically increases with $r$
and becomes $1$ at $r=d$.
Moreover, when $r$ is fixed,
fidelity monotonically decreases as $d$ increases
and becomes $0$ in the limit $d\to\infty$.

%%%%%%%%%%%%%%%%%%%%%%%%%%%%%%%%%%%%%%%%%%%%%%%%%%%%%%%%%%%%%%%%%%%
In terms of the density operator of the system,
the measurement changes the maximally mixed state in $d$ dimensions,
$\hat{\rho}=\sum_a p(a) \ket{\psi(a)}\bra{\psi(a)}=\hat{I}/d$,
into that in $r$ dimensions,
decreasing the von Neumann entropy of the system by
$\log_2d-\log_2r=\log_2(d/r)$.
However, the information gain (\ref{eq:information})
is less than $\log_2(d/r)$
because of our formulation of information resource~\cite{NieChu00},
i.e. a set of predefined states with Shannon entropy
rather than a density operator with von Neumann entropy.
Within this formulation,
the second term in Eq.~(\ref{eq:information})
comes from the indistinguishability of non-orthogonal quantum states.
To see this,
consider the orthonormal basis $\{\ket{k}\}$
with $k=1,2,\ldots,d$ in Eq.~(\ref{eq:basis})
as the set of predefined states,
instead of all possible pure states $\{\ket{\psi(a)}\}$.
In this distinguishable case,
the information gain is equal to just
the decrease in the von Neumann entropy $\log_2(d/r)$.
Therefore, the reduced information gain (\ref{eq:information})
is due to the indistinguishability of predefined states.
In other words,
quantum measurement with no \textit{a priori}
information about the state of the system
is not optimal
as quantum communication between the system and observer,
since its encoding and decoding procedure
using all possible pure states
suffers information loss
by the indistinguishability of states.
%%%%%%%%%%%%%%%%%%%%%%%%%%%%%%%%%%%%%%%%%%%%%%%%%%%%%%%%%%%%%%%%%%%

\section{\label{sec:tradeoff}Tradeoff and Efficiency}
\begin{figure}
\begin{center}
\includegraphics[scale=0.6]{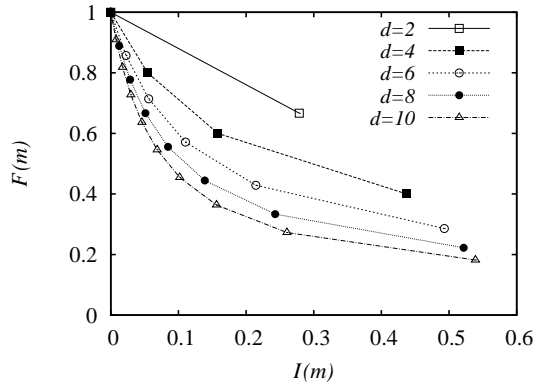}
\end{center}
\caption{\label{fig3}
Fidelity $F(m)$
as a function of information gain $I(m)$ for $d=2,4,6,8,10$.}
\end{figure}

From the explicit formulae for the information gain (\ref{eq:information})
and fidelity (\ref{eq:fidelity}),
we find a tradeoff between information and fidelity
in projective measurements.
Figure~\ref{fig3} shows the fidelity $F(m)$
as a function of the information gain $I(m)$ for $d=2,4,6,8,10$.
As the measurement provides more information about
the state of a system,
the process of measurement changes the state to a greater extent,
as shown in Fig.~\ref{fig3}.
It should be emphasized
that this tradeoff is at a single outcome level
in the sense that there is no average over outcome.

\begin{figure}
\begin{center}
\includegraphics[scale=0.6]{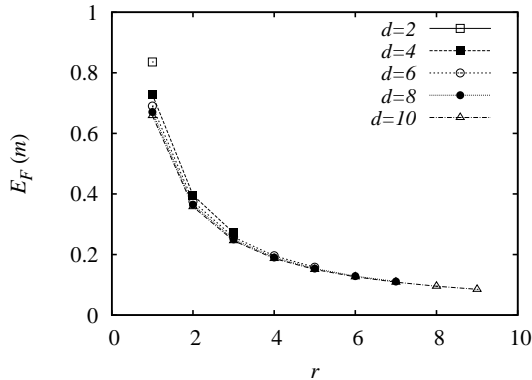}
\end{center}
\caption{\label{fig4}
Efficiency of measurement $E_F(m)$
as a function of rank $r$ for $d=2,4,6,8,10$.}
\end{figure}

In addition, another relationship
between information gain and state change
can be shown by defining the efficiency of measurement as
the ratio of the information gain
to the fidelity loss~\cite{Terash10,Terash11},
\begin{equation}
    E_F(m)\equiv \frac{I(m)}{1-F(m)}.
\end{equation}
Figure~\ref{fig4} shows the efficiency of measurement, $E_F(m)$,
as a function of rank $r$ for $d=2,4,6,8,10$,
although it is ill-defined at $r=d$ because of $I(m)=1-F(m)=0$.
The efficiency is a monotonically decreasing function for each $d$
and has a maximal value $3[1-1/(2\ln2)]$ at $r=1$ in $d=2$.
This means that among the various projective measurements,
a projective measurement on a two-level system or qubit
is the most efficient with respect to fidelity.
Nevertheless, it is the least efficient among single-qubit measurements,
as discussed in Ref.~\cite{Terash11}.

\section{\label{sec:conclude}Conclusion}
We calculated the information gain and fidelity
of a projective measurement on a system
where the pre-measurement state was assumed
to be in a completely unknown state.
They are expressed as functions of the dimensions $d$ of the
Hilbert space associated with the system
and rank $r$ of the projection operator associated with the measurement,
as in Eqs.~(\ref{eq:information}) and (\ref{eq:fidelity}).
These results show a tradeoff
between information and fidelity at the level of a single outcome
without averaging all outcomes, as shown in Fig.~\ref{fig3}.
We also discussed the efficiency of the measurement by using
the ratio of information gain to fidelity loss.
In terms of this efficiency,
a projective measurement on a two-level system or qubit is
the most efficient among the various projective measurements.

Although here we have considered only projective measurements,
there are many measurements that are not projective, e.g.,
photodetection processes in photon counting~\cite{Terash10}.
Such measurements can be less informative but more reversible
than projective measurements.
However, in general measurements on a $d$-level system,
it would be difficult to find tradeoffs among
information gain, fidelity, and physical reversibility
because they are all functions of $d-1$ parameters~\cite{Terash11}.
To find the tradeoffs,
our present results suggest
some special points such as the endpoints of boundary curves
in the tradeoffs,
since projective measurements are the most informative but
the least reversible.

\section*{Acknowledgments}
This study was supported by a Grant-in-Aid
for Scientific Research (Grant No.~20740230) from
Japan Society for the Promotion of Science.

%\bibliographystyle{prsty}
%\bibliography{art2,art3,bk,paper}
%\end{document}

\end{document}